\documentclass{ws-p8-50x6-00}

\newcommand{\ba}[1]{\begin{eqnarray} \label{(#1)}}
\newcommand{\ea}{\end{eqnarray}}

\newcommand{\AmS}{{\protect\the\textfont2
  A\kern-.1667em\lower.5ex\hbox{M}\kern-.125emS}}

\def\be{\begin{equation}}
\def\ee{\end{equation}}
\def\bea{\begin{eqnarray}}
\def\eea{\end{eqnarray}}
\def \gsim {~\mbox{${}^> \hspace*{-9pt} _\sim$}~}

\begin{document} 

%\begin{center}
%{\bf 
\title{Information about the neutrino mass matrix from Double Beta Decay}

\author{H.V. Klapdor-Kleingrothaus}

\address{Max-Planck-Institut f\"ur Kernphysik,\\ 
P.O. Box 10 39 80, D-69029 Heidelberg, Germany\\ 
Spokesman of HEIDELBERG-MOSCOW and GENIUS Collaborations\\
E-mail: klapdor@gustav.mpi-hd.mpg,\\
 Home-page: http://mpi-hd.mpg.de.non\_acc/}
%\end{center}

\maketitle

%%%%%%%%%%%%%%%%%%%%%%%%%% ABSTRACT %%%%%%%%%%%%%%%%%%%%%%%%%%%%%%

\abstracts{
	Double beta decay is indispensable to solve the question of 
	the neutrino mass matrix {\it together} with $\nu$ oscillation 
	experiments. 
	The most sensitive experiment since eight years --- 
	the HEIDELBERG-MOSCOW experiment in Gran-Sasso --- already now, 
	with the experimental limit of 
$\langle m_\nu \rangle < 0.26$~eV excludes degenerate $\nu$ mass scenarios 
	 allowing neutrinos as hot dark matter in the universe for the 
	 small angle MSW solution of the solar neutrino problem. 
	 It probes cosmological models including hot dark matter 
	 already now on the level of future satellite 
	 experiments MAP and PLANCK. 
	 It further probes many topics of beyond Standard Model 
%SM 
	 physics at the TeV scale. 
	 Future experiments should give access to the multi-TeV 
	 range and complement on many ways the search for new physics 
	 at future colliders like LHC and NLC. 
	 For neutrino physics some of them (GENIUS) will allow to test 
	 almost {\it all} neutrino mass scenarios allowed by 
	 the present neutrino oscillation experiments. 
	 A GENIUS Test Facility has just been funded and will 
	 come into operation by end of 2001.} 

%%%%%%%%%%%%%%%%%%%%%%%%%% end ABSTRACT %%%%%%%%%%%%%%%%%%%%%%%%%%%%%%

%%%%%%%%%%%%%%%%%%%%%%%%%% Introduction %%%%%%%%%%%%%%%%%%%%%%%%%
\section{Introduction}
	Recently atmospheric and solar neutrino oscillation experiments 
	have shown that neutrinos are massive. This is the first 
	indication of beyond standard model physics. The absolute 
	neutrino mass scale is, however, still unknown, 
	and only neutrino oscillations and neutrinoless double beta decay 
	{\it together} can solve this problem (see, e.g. 
\cite{KKPS,KKPS-01,KK60Y}).

	In this paper we will discuss the contribution, that can be 
	given by present and future $0\nu\beta\beta$ experiments to this 
	important question of particle physics. We shall, in section 2, 
	discuss the expectations for the observable of neutrinoless double 
	beta decay, the effective neutrino mass $\langle m_\nu \rangle$, 
	from the most recent $\nu$ oscillation experiments, 
	which tells us the required sensitivity for future $0\nu\beta\beta$ 
	experiments. In section 3 we shall discuss the present status 
	and future potential of $0\nu\beta\beta$ experiments. 
	It will be shown, that if by exploiting the potential of 
	$0\nu\beta\beta$ decay to its ultimate experimental limit, it will 
	be possible to test practically 
	{\it all} neutrino mass scenarios allowed by the present neutrino 
	oscillation experiments (except for one, the hierarchical 
	LOW solution). 
%%%%%%%%%%%%%%%%%%%%%%%%% End Introduction %%%%%%%%%%%%%%%%%%%%%%%%%%

%%%%%%%%%%%%%%%%%%%%%%%%%% Section 1 %%%%%%%%%%%%%%%%%%%%%%%%%% 

\section{\boldmath Allowed ranges of 
$\langle m \rangle$ by $\nu$ oscillation experiments}
	 After the recent results from Superkamiokande (e.g. see 
\cite{Suz00,Val01}), the prospects for a positive signal in $0\nu\beta\beta$ 
	  decay have become more promising. 
	  The observable of double beta decay 
$\langle m \rangle =
|\sum U^2_{ei}m^{}_i| = |m^{(1)}_{ee}| 
		      + e^{i\phi_2} |m^{(2)}_{ee}| 
		      + e^{i\phi_3} |m^{(3)}_{ee}|
$
	  with $U^{}_{ei}$ 
	  denoting elements of the neutrino mixing matrix, 
	  $m_i$ neutrino mass eigenstates, and $\phi_i$  relative Majorana 
	  CP phases, can be written in terms of oscillation parameters 
\cite{KKPS,KKPS-01} 
\begin{eqnarray}
\label{1}
|m^{(1)}_{ee}| &=& |U^{}_{e1}|^2 m^{}_1,\\
\label{2}
|m^{(2)}_{ee}| &=& |U^{}_{e2}|^2 \sqrt{\Delta m^2_{21} + m^{2}_1},\\
\label{3}
|m^{(3)}_{ee}| &=& |U^{}_{e3}|^2 \sqrt{\Delta m^2_{32} 
				 + \Delta m^2_{21} + m^{2}_1}.
\end{eqnarray}

	The effective mass $\langle m \rangle$ is related with the 
	half-life for $0\nu\beta\beta$ decay via 
$\left(T^{0\nu}_{1/2}\right)^{-1}\sim \langle m_\nu \rangle^2$, 
        and for the limit on  $T^{0\nu}_{1/2}$
	deducible in an experiment we have 
	$T^{0\nu}_{1/2} \sim a \sqrt{\frac{Mt}{\Delta E B}}$.
	Here $a$ is the isotopical abundance of the $\beta\beta$ emitter;
	$M$ is the active detector mass; 
	$t$ is the measuring time; 
	$\Delta E$ is the energy resolution; 
	$B$ is the background count rate. 

%%%%%%%%%%%%%%%%%%%%%%%%% begin. fig. 1 %%%%%%%%%%%%%%%%%%%%%%%%
%
 \vspace{-0.5cm}
\begin{figure}[ht]
\vspace{9pt}
\centering{
\includegraphics*[scale=0.55]
{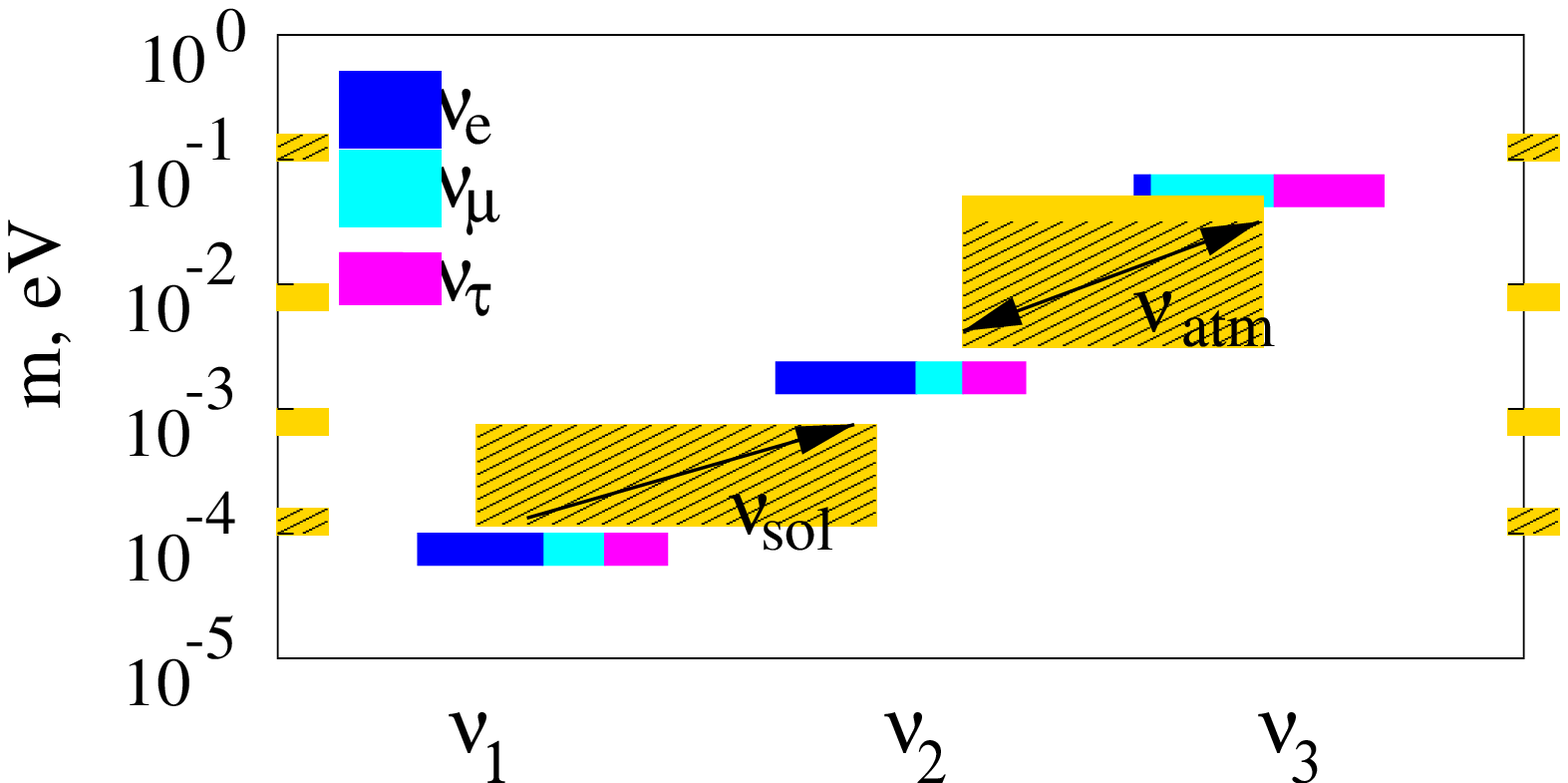}}
\vspace{-0.5cm}
\caption[]{%Figure 1. 
       Neutrino masses and mixings in the scheme with mass hierarchy. 
       Coloured bars correspond to flavor admixtures in the mass 
       eigenstates $\nu_1, \nu_2, \nu_3$. 
       The quantity $\langle m \rangle$
	 is determined by the dark blue bars denoting 
	 the admixture of the electron neutrino $U_{ei}$.
\label{fig:Hierarchi-NuMass}}
\end{figure}
% 
%%%%%%%%%%%%%%%%%%%%%% end fig. 1 %%%%%%%%%%%%%%%%%%%%%%%%%%%%%%%
\vspace{-0.5cm}	
	Neutrino oscillation experiments fix or restrict some of the 
	parameters in 
(1)--(3), e.g. in the case of normal hierarchy solar neutrino 
	  experiments yield 
	  $\Delta m^2_{21}$, 
	  $|U_{e1}|^2 = \cos^2\theta_{\odot}$ 
	  and
	  $|U_{e2}|^2 = \sin^2\theta_{\odot}$. 
	  Atmospheric neutrinos fix  
	  $\Delta m^2_{32}$, 
	  and experiments like CHOOZ, looking for $\nu_e$ 
	  disappearance restrict $|U_{e3}|^2$. 
	  The phases $\phi_i$  and the mass of the lightest neutrino, 
	  $m_1$ are free parameters. 
	  The expectations for 
$\langle m \rangle$ 
	  from oscillation experiments in different neutrino mass scenarios 
	  have been carefully analyzed in% 
\cite{KKPS,KKPS-01}. In sections 2.1 to 2.3 we give some examples.

%%%%%%%%%%%%%%%%%%%%%%%%%% end section 1 %%%%%%%%%%%%%%%%%%%%%%%%%% 

%%%%%%%%%%%%%%%%%%%%% Beg. fig. 2 %%%%%%%%%%%%%%%%%%%%%%

%\clearpage
\begin{figure}[ht]
%\vspace{9pt}
\centering{\includegraphics*[scale=0.40]{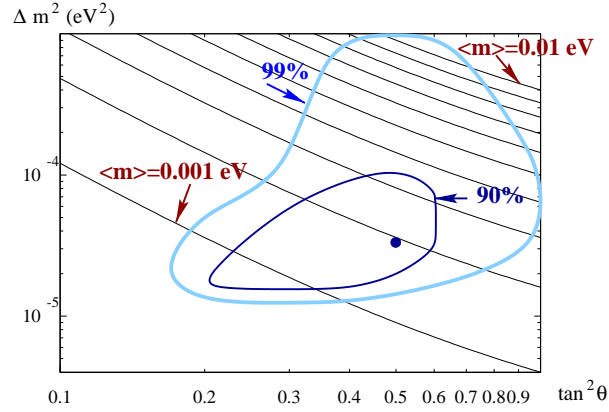}}
%\vspace{-0.9cm}
\caption[]{%Figure 2. 
       Double beta decay observable
$\langle m \rangle$
       and oscillation parameters in the case of the MSW 
       large mixing angle solution of the solar neutrino deficit, 
       where the dominant contribution to $\langle m \rangle$ comes 
       from  the  second  state. 
       Shown are  lines of constant $\langle m \rangle$,  
       the lowest  line corresponding to 
       $\langle m_\nu \rangle = 0.001$~eV, 
       the upper line to 0.01~eV. 
       The inner and outer closed line show the regions allowed 
       by present solar neutrino experiments with 
	90\% C.L. and 99\% C.L., 
       respectively. 
       Double beta decay with sufficient sensitivity could check the 
       LMA MSW solution. 
       Complementary information could be obtained from the search for a 
       day-night effect and spectral distortions in future solar 
       neutrino experiments as well as a disappearance signal in KAMLAND.
\label{fig:Dark2}
}
\vspace{.5cm}
\end{figure}

%%%%%%%%%%%%%%%%%%%%% end fig. 2 %%%%%%%%%%%%%%%%%%%%%%

%%%%%%%%%%%%%%%%%%%%%%%%%% section Hierarch sp. %%%%%%%%%%%%%%%%%%%%%%%%%% 
\vspace{-1.2cm}
\subsection{\boldmath Hierarchical spectrum $(m_1 \ll m_2 \ll  m_3)$}
         In hierarchical spectra
(Fig.~\ref{fig:Hierarchi-NuMass}), 
	motivated by analogies with the quark sector and the simplest 
	  see-saw models, the main contribution comes from 
	  $m_2$ or $m_3$. 
	  For the large mixing angle (LMA) MSW solution which is favored 
	  at present for the solar neutrino problem (see 
\cite{Suz00}), the contribution of $m_2$ becomes dominant in the expression 
       for $\langle m \rangle$, and  
\begin{equation}
\langle m \rangle \simeq m^{(2)}_{ee} 
	= \frac{\tan^2\theta}{1+\tan^2 \theta}\sqrt{\Delta m^2_{\odot}}.
\end{equation}
	In the region allowed at 90\% C.L. by Superkamiokande according to 
\cite{Val01}, 
	the prediction for $\langle m \rangle$, becomes        
\begin{equation}
\langle m \rangle =(1\div 3) \cdot 10^{-3} {\rm eV}.
\end{equation}
	The prediction extends to 
	$\langle m \rangle = 10^{-2}$ eV in the 99\% C.L. range 
(Fig. ~\ref{fig:Dark2}).

%%%%%%%%%%%%%%%%%%%%%%%%%% end Hierarch sp.%%%%%%%%%%%%%%%%%%%%%%%%%%%%%

%%%%%%%%%%%%%%%%%%%%% Beg. fig. 3 a, b %%%%%%%%%%%%%%%%%%%%%%
\vspace{-.5cm}
\begin{figure}[t]
%\vspace{9pt}
\centering{
\includegraphics[scale=0.37]
{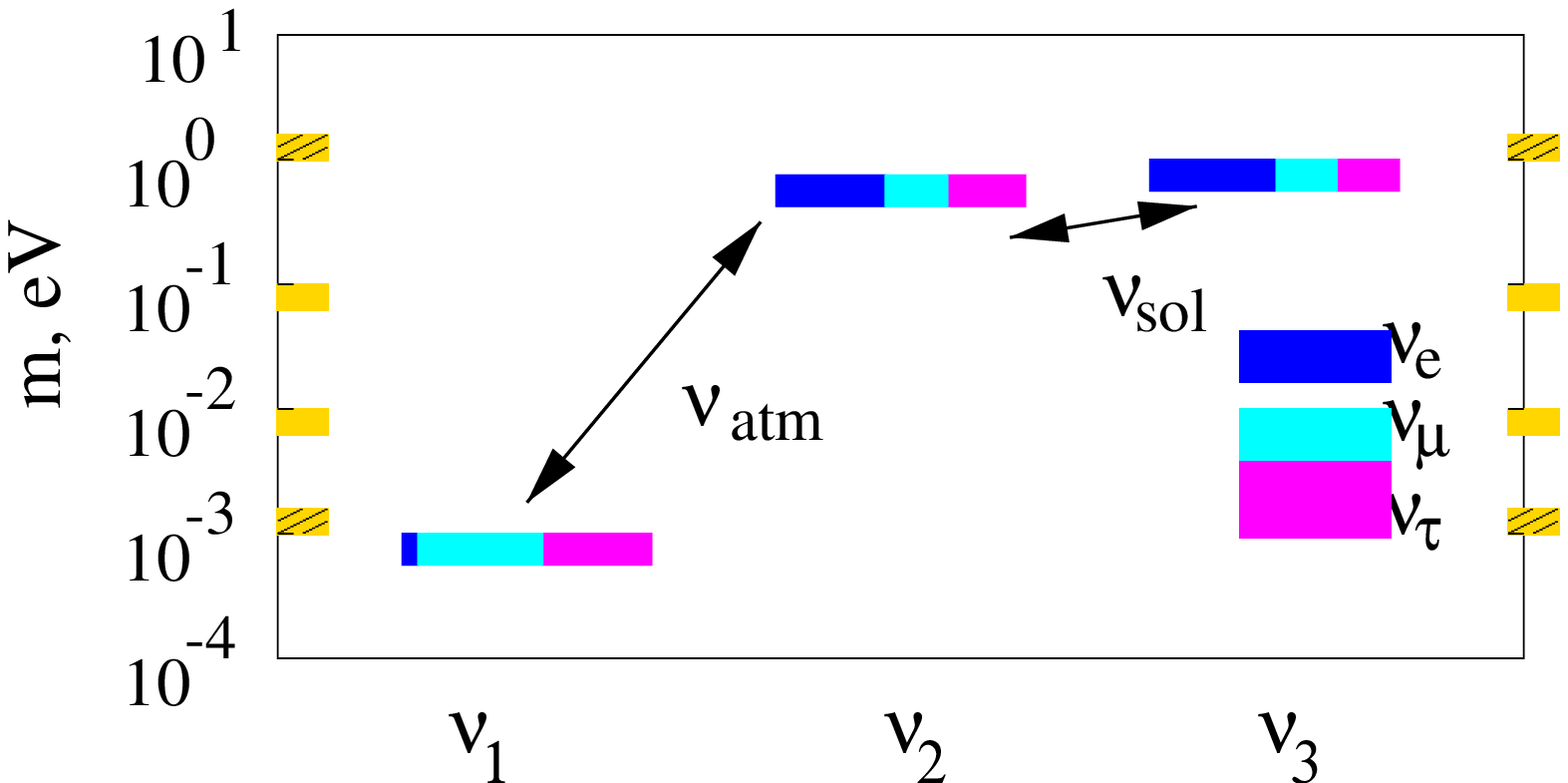} \hfill
\includegraphics[scale=0.37]
{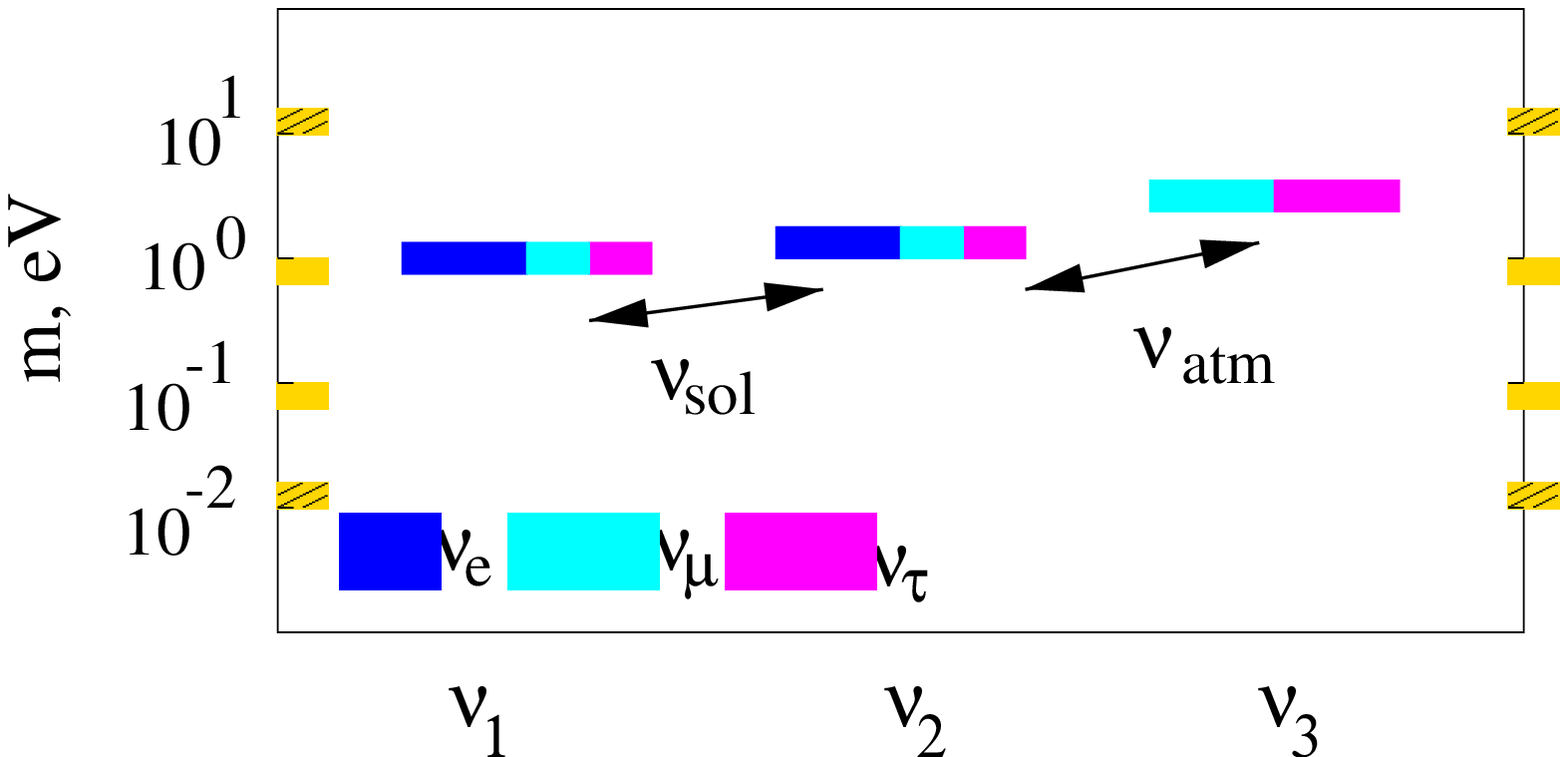}
}
\caption[]{%Figure 3. 
       \underline{Left:} Neutrino masses and mixing in the inverse 
	hierarchy scenario. \underline{Right:} Neutrino masses and mixings 
	in the degenerate scheme.
\label{fig:INVERSE-NuMass}
\label{fig:Degener-NuMass}}
\end{figure}

%%%%%%%%%%%%%%%%%%%% end fig. 3 a, b %%%%%%%%%%%%%%%%%%%%%%

%%%%%%%%%%%%%%%%% section  Invers. Hierarch. %%%%%%%%%%%%%%%%%%%%%%%%%%%%%%
\vspace{-1.cm}
\subsection{\boldmath Inverse Hierarchy $(m_3 \approx m_2 \gg  m_1)$}
           In inverse hierarchy scenarios 
(Fig.~\ref{fig:INVERSE-NuMass})   
	the heaviest state with mass $m_3$ is mainly the electron 
	   neutrino, its mass being determined by atmospheric neutrinos, 
$m_3 \simeq \sqrt{\Delta m^2_{\rm atm}}$.
	   For the LMA MSW solution one finds 
\cite{KKPS-01}
\begin{equation}
\langle m \rangle 
= (1\div 7) \cdot 10^{-2} {\rm eV}.
\end{equation}

%%%%%%%%%%%%%%%%% end section  Invers. Hierarch.%%%%%%%%%%%%%%%%%%%%%%%%%%%%

%%%%%%%%%%%%%%%%% begin. fig.4 *************************
\vspace{-.5cm}
\begin{figure}[t]
%\vspace{9pt}
\centering{
\includegraphics*[scale=0.34, angle=-90]
{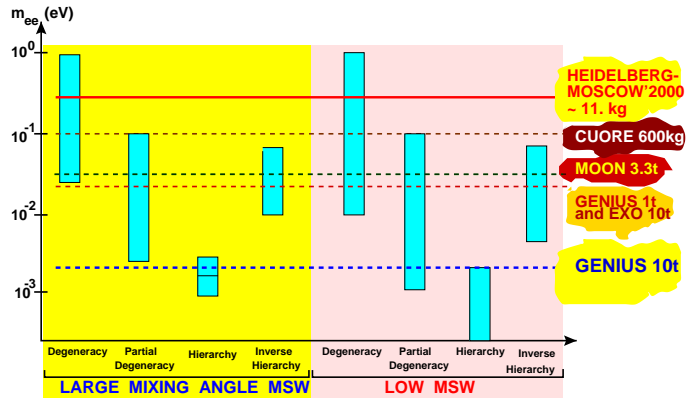}}
\caption[]{%Figure 5. 
       	Summary of values for $m_{ee} =\langle m \rangle$
	expected from neutrino oscillation experiments 
	(status NEUTRINO2000), in the different schemes discussed in 
	this paper. 
	For a more general analysis see 
\cite{KKPS}. 
	The expectations are compared with the recent neutrino mass 
	limits obtained from the HEIDELBERG-MOSCOW 
\cite{KK01,AnnRepGrSs00}
	, experiment as well as the expected sensitivities for the CUORE 
\cite{CUORE98},  
MOON \cite{Ej00}, 
EXO \cite{EXO00} proposals and the 1 ton and 10 ton proposal of 
      GENIUS 
\cite{KK-Bey97,GEN-prop}.
\label{fig:Jahr00-Sum-difSchemNeutr}}
\end{figure}
%%%%%%%%%%%%%%%%% end fig.4 *************************

%%%%%%%%%%%%%%%%%  section   Degenerate spectrum %%%%%%%%%%%%%%%%%%%%%%%%%%%%

\subsection{\boldmath Degenerate spectrum $(m_1 \simeq m_2 \simeq m_3 
	\gsim	 0.1~$eV)}
%%%%%%%%%
        In degenerate scenarios (fig. \ref{fig:Degener-NuMass}) the 
	contribution of $m_3$ is strongly restricted by CHOOZ.  
	The main contributions come from $m_1$ and $m_2$, depending on 
	their admixture to the electron flavors, which is determined 
	by the solar neutrino solution. We find 
\cite{KKPS-01}
\begin{equation}
m_{\min} < \langle m \rangle < m_1 \qquad 
\mbox{with} \qquad 
\langle m_{\min}\rangle = 
	(\cos^2\theta_{\odot} -\sin^2\theta_{\odot})\, m^{}_1.
\end{equation}

%\vspace{-.8cm}
       	This leads for the LMA solution to 
$\langle m \rangle = (0.25\div 1)\cdot m_1$, 
	 the allowed range corresponding to possible values of 
	 the unknown Majorana CP-phases.

	 After these examples we give a summary of our analysis 
\cite{KKPS,KKPS-01} 
	of the $\langle m \rangle $ allowed by $\nu$ oscillation 
      experiments for neutrino mass models in the presently 
      favored scenarios, 
in Fig.~\ref{fig:Jahr00-Sum-difSchemNeutr}. 
	  The size of the bars corresponds to the uncertainty in 
	  mixing angles and the unknown Majorana CP-phases.

%%%%%%%%%%%%%%%%% end section  Degener.  %%%%%%%%%%%%%%%%%%%%%%%%%%%%%%%%%%%

%%%%%%%%%%%%%%%%%%%% Beg. Fig. 5 *************************
%\newpage
\vspace{-.3cm}
\begin{figure}[b]
%\vspace{9pt}
\centering{
\includegraphics*[scale=0.35, angle=-90 ]
{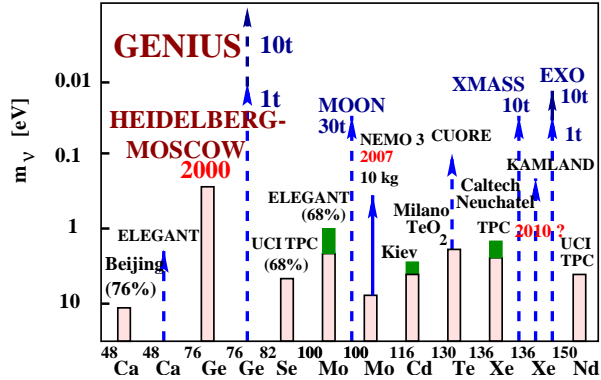}}

%\vspace{-.5cm}
\caption[]{%Figure 5. 
       Present situation, 2000, and expectation for the future, 
       of the most promising $\beta\beta$ experiments. 
       Light parts of the bars: present status; dark parts: 
       expectation for running experiments; solid and dashed lines: 
       experiments under construction or proposed experiments, respectively. 
       For references see 
\protect\cite{KK60Y,KK-LowNu2,LowNu2}.
\label{fig:Now4-gist-mass}}
\end{figure}

%%%%%%%%%%%%%%%%%%%% end Fig. 5 *************************

%%%%%%%%%%%%%%%%%%%% beg. Fig. 6 *************************

\begin{figure}[t]
%\vspace{9pt}
\centering{
\includegraphics*[scale=0.35, angle=-90]
{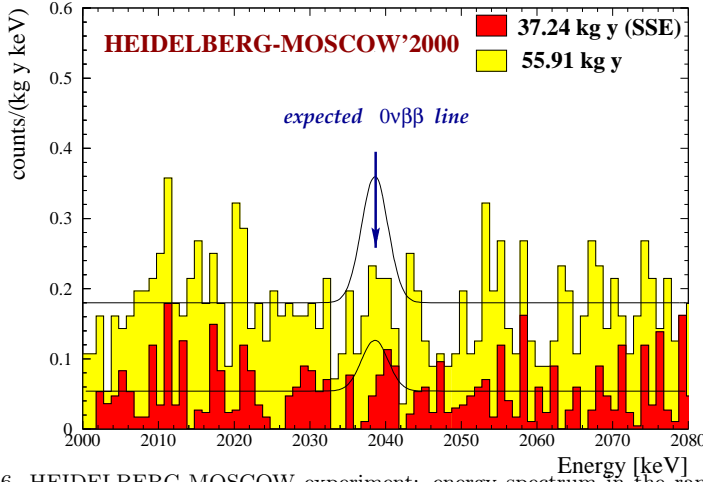}}

\vspace{-.5cm}
\caption[]{%Figure 6. 
       HEIDELBERG-MOSCOW experiment: energy spectrum in the range 
       between 2000 keV and 2080 keV, where the peak from neutrinoless 
       double beta decay is expected. 
       The open histogram denotes the overall sum spectrum without 
       PSA after 55.9 kg y of measurement (since 1992). 
       The filled histogram corresponds to the SSE data after 37.2 kg y. 
       Shown are also the excluded (90\% C.L.) peak areas from the 
	two spectra.
\label{fig:Spectr-37-24kgy}}
\end{figure}

%%%%%%%%%%%%%%%%%%%% end Fig. 6 *************************

%%%%%%%%%%%%%%%%% section 3. Status %%%%%%%%%%%%%%%%%%%%%%%%%%%%%%%%%%%%%%

\section{\boldmath Status of $\beta\beta$ Experiments}
	The status of present double beta experiments is shown in 
Fig.~\ref{fig:Now4-gist-mass}	
	and is extensively discussed in 
\cite{KK60Y}.	
	The HEIDELBERG-MOSCOW experiment using the largest source strength 
	of 11 kg of enriched $^{76}$Ge in form of five HP Ge-detectors in 
	the Gran-Sasso underground laboratory 
\cite{KK60Y,KK-StProc00}, 
	yields after a time of 37.2~kg$\cdot$y of measurement
(Fig.~\ref{fig:Spectr-37-24kgy})  a half-life limit of 
\cite{AnnRepGrSs00,AnnRepMPI00} 
$$
T^{o\nu}_{1/2} > 2.1(3.5)\cdot 10^{25}\ {\rm y}, \quad 
	       90\%~(68\%)~{\rm C.L.}
$$
	and a limit for the effective neutrino mass of 
$$
\langle m\rangle 
	< 0.34(0.26)\ {\rm eV}, \quad   
		 90\%~(68\%)~{\rm C.L.}
$$
	This sensitivity just starts to probe some (degenerate) neutrino 
	mass models (see 
Fig.~\ref{fig:Jahr00-Sum-difSchemNeutr}). 
	 In degenerate models from the experimental limit on 
$\langle m\rangle $
	 we can conclude an upper bound on the mass scale of the heaviest 
	 neutrino. 
	 For the LMA solar solution we obtain from 
(7)	 $m_{1,2,3}< 1.1\,$eV implying 
	 $\sum m_i < 3.2\,$eV. 
	 This first number is sharper than what has recently been deduced 
	 from single beta decay of tritium 
	 ($m < 2.2$~eV 
\cite{Weinh-Neu00}),  
	and the second is sharper than the limit of 
	 $\sum m_i < 5.5\,$eV 
	 still compatible with most recent fits of 
	 Cosmic Microwave Background Radiation 
	 and Large Scale Structure data (see, e.g. 
\cite{Teg00}).

%%%%%%%%%%%%%%%%%%%% beg. Fig. 7 *************************

\begin{figure}[t]
\vspace{9pt}
\centering{
\includegraphics*[scale=0.35, angle=-90]
{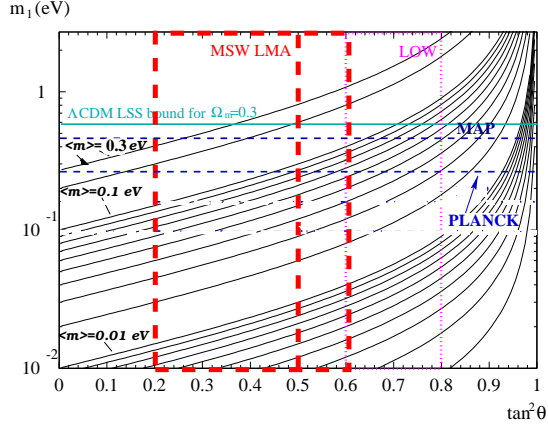}}
\caption[]{%Figure 7. 
       Double beta decay observable 
$\langle m\rangle$
	 and oscillation parameters: 
	 The case for degenerate neutrinos. 
	 Plotted on the axes are the overall scale of neutrino masses 
	 $m_0$ and mixing $\tan^2\, 2\theta^{}_{12}$. 
	 Also shown is a cosmological bound deduced from a fit of 
	 CMB and large scale structure 
\cite{Lop} 
	and the expected sensitivity of the satellite experiments 
	 MAP and PLANCK. 
	 The present limit from tritium $\beta$ decay of 2.2 eV 
\cite{Weinh-Neu00} 
	would lie near the top of the figure. 
     The range of 
$\langle m\rangle$
	 investigated at present by the HEIDELBERG-MOSCOW experiment is, 
	 in the case of small solar neutrino mixing already in the 
	 range to be explored by MAP and PLANCK 
\cite{Lop}.
\label{fig:Dark3}}
\end{figure}
%\vspace{-0.7cm}
%%%%%%%%%%%%%%%%%%%% end Fig. 7 *************************

	The result has found a large resonance, and it has been shown that 
	it excludes for example the small angle MSW solution of the solar 
	neutrino problem in degenerate scenarios, if neutrinos are 
	considered as hot dark matter in the universe% 
\cite{Glash00,Min00,Yas-Bey99,ElLol99}. 
	This conclusion has been made, {\it before} the SMA MSW solution 
	has been disfavored by the SUPERKAMIOKANDE collaboration in June 2000. 
Figure \ref{fig:Dark3} shows that the present sensitivity 
	 probes cosmological models including hot dark matter already 
	 now on a level of future satellite experiments MAP and PLANCK. 
	It starts to become interesting also in connection with 'Z-burst' 
	models recently discussed as explanation for super-high energy 
	cosmic ray events beyond the GZK cutoff energy% 
\cite{Wei82,Wei-Bey99,Far99,WeiPaes01}.

	 The HEIDELBERG-MOSCOW experiment, 
%using the world's largest source strength, 
	 yields now since eight years already the by far sharpest 
	 limits worldwide. 
	 If future searches will show that 
$\langle m\rangle > 0.1$~eV, then the three-$\nu$ mass schemes, 
	 which will survive, are those with $\nu$ mass degeneracy, and in
	 4-neutrino schemes, those with inverse mass hierarchy 
(Fig. \ref{fig:Jahr00-Sum-difSchemNeutr} and see also ref.%
\cite{KKPS}).

      It has been discussed in detail earlier (see e.g.% 
\cite{KK60Y,KK-Bey97,KK-Neutr98,KK-NOW00}
	), that of present generation experiments no one has a 
	   potential to probe 
$\langle m\rangle$
     below the present HEIDELBERG-MOSCOW level (see 
Fig.~\ref{fig:Now4-gist-mass}).

	A second experiment using enriched $^{76}$Ge, IGEX, has stopped 
	operation by end of 1999 
\cite{IGEX00}. 
	This experiment already started in 1992 with 2.1 kg of 
	$^{76}$Ge 
\cite{IGEX93} and operated in 1995 already 8 
     kg of $^{76}$Ge 
\cite{IGEX96}.	In 1999 they published a measuring time of 5.7 kg y 
	(less than one year of full operation) 
\cite{IGEX97,IGEX99}, and in autumn 99 of about 9 kg y 
\cite{IGEX-taup99} 
	 (less than one quarter of the HEIDELBERG-MOSCOW significance) 
	 and an optimistic value for 
$\langle m\rangle$, 
	 using a method criticized.
	 The Milano cryogenic experiment using TeO$_2$ bolometers 
	 improved their values for the 
$\langle m_\nu\rangle$ 
	 from $\beta\beta$ decay of $^{130}$Te, from 5.3 eV in 1994%  
\cite{Ales94} to 1.8 eV in 2000%  
\cite{Ales00}, and according to%  
\cite{CUORE01} to 0.9 eV in early 2001.
	Also CUORICINO (with 45 kg of detectors) scheduled for starting in 
	autumn 2001%  
\cite{CUORE01} will hardly reach the HEIDELBERG-MOSCOW limit 
	(see also discussion in 
\cite{Bell00}).
	NEMO-III, originally aiming at a sensitivity of 0.1 eV, 
	reduced their goals recently to $0.3\div0.7$~eV (see% 
\cite{NEMO-Neutr00}
	), (which is more consistent with estimates given by%  
\cite{Tret95}
	), to be reached in 6 years from starting of running, 
       foreseen for the year 2002.

	A possibility to probe $\langle m \rangle$ down to $\sim 0.1$ eV 
	(90\% c.l.) exists with the GENIUS Test Facility%  
\cite{AnnRepMPI00} (see below),
	which should reduce the background by a factor of 30  compared to 
	the HEIDELBERG-MOSCOW experiment, and thus 
	could reach a half-life limit of $1.5 \cdot {10}^{26}$ y.

%%%%%%%%%%%%%%%%% end sect. 3  STATUS %%%%%%%%%%%%%%%%%%%%%%

%%%%%%%%%%%%%%%%% sect. 4 Future %%%%%%%%%%%%%%%%%%%%%%

\section{\boldmath Future of $\beta\beta$ Experiments}
		   To extend the present sensitivity of 
		   $\beta\beta$ experiments below a limit of 0.1 eV, 
		   requires completely new experimental approaches, 
		   as discussed extensively in% 
\cite{KK60Y,KK-Bey97,GEN-prop,KK-Neutr98}.

	Figure \ref{fig:Jahr00-Sum-difSchemNeutr} 
	shows that an improvement of the sensitivity down to 
$\langle m\rangle\sim 10^{-3}$~eV 
	 is required to probe all neutrino mass scenarios allowed by 
	 present neutrino oscillation experiments% 
\cite{KK-Bey97,KKPS}. 
	With this result of $\nu$ oscillation experiments nature seems 
	to be generous to us since such a sensitivity seems to be 
	achievable in future $\beta\beta$ experiments, 
	if this method is exploited to its ultimate limit, 
	as by the GENIUS project% 
\cite{KK60Y,KK-Bey97,GEN-prop,KK-Neutr98,KK-WEIN98,KK-InJModPh98,KK-J-PhysG98,KK-LeptBar98,KK-SprTracts00,KK-LowNu2}.

%%%%%%%%%%%%%%%%% subsect. 4.1 /////////////////////////

\subsection{\boldmath GENIUS, Double Beta Decay and the Light Majorana 
		      Neutrino Mass}
		    
%\vspace{.3cm}
	With the era of the HEIDELBERG-MOSCOW experiment 
		    which will remain the most sensitive experiment 
		    for the next years, the time of the small smart 
		    experiments is over.

            The requirements in sensitivity for future experiments to play 
	    a decisive role in the solution of the structure of the 
	    neutrino mass matrix can be read from  
Fig.~\ref{fig:Jahr00-Sum-difSchemNeutr}.

	     To reach the required level of sensitivity $\beta\beta$ 
	     experiments have to become large. 
	     On the other hand source strengths of up to 10 tons of 
	     enriched material touch the world production limits. 
	     At the same time the background has to be reduced by a 
	     factor of 1000 and more compared to that 
	     of the HEIDELBERG-MOSCOW experiment.

Table~1 lists some key numbers for GENIUS,% 
\cite{KK-Bey97,GEN-prop,KK-J-PhysG98} 
	     which was the 
	     first proposal for a third generation double beta experiment, 
	     and which may be {\em the only}\ project, which will be able to 
	     test {\em all}\ neutrino mass scenarios, and of some other 
	     proposals made {\em after}\ the GENIUS proposal. 
	     The potential of some of them is shown also in 
Fig.~\ref{fig:Jahr00-Sum-difSchemNeutr} and 
Fig.~\ref{fig:Now4-gist-mass}.	     
	It is seen that not all of these proposals fully cover 
	     the region to be probed. 
	Among them is also the recently presented MAJORANA project 
\cite{MAJOR-WIPP00}, 
	which does not really apply any new strategy for background reduction.
	For more recent information on XMASS, EXO, MOON experiments see 
	the contributions of Y. Suzuki, G. Gratta and H. Ejiri in these  
	Proceedings% 
\cite{LowNu2}.
	     The CAMEO project% 
\cite{Bell00} 
	will have to work on {\em very}\ long time scales, also 
	     since it has to wait the end of the BOREXINO 
	     solar neutrino experiment.
	     CUORE% 
\cite{CUORE-LeptBar98} 
	has, with the complexity of cryogenic techniques, 
	   still to overcome serious problems of background to enter 
	   into interesting regions of
$\langle m_\nu\rangle$.
	 EXO% 
\cite{EXO00} 
	needs still very extensive research and development to probe 
	 the applicability of the proposed detection method.

	 In the GENIUS project a reduction by a factor of more than 1000 
	 down to a background level of 0.1 events/tonne y keV 
	 in the range of $0\nu\beta\beta$ decay is reached by removing all 
	 material close to the detectors, and by using naked Germanium 
	 detectors in a large tank of liquid nitrogen. 
	 It has been shown that the detectors show excellent 
	 performance under such conditions% 
\cite{GEN-prop,KK-Bey97}.

	For technical questions and extensive Monte Carlo simulations of 
	the GENIUS project for its application in double beta decay 
	we refer to% 
\cite{GEN-prop,KK-J-PhysG98}.

%%%%%%%%%%%%%%%%% end subsect. 4.1 /////////////////////////

%%%%%%%%%%%%%%%%% subsect. 4.2 /////////////////////////
\vspace{-.45cm}
\subsection{GENIUS and Other Beyond Standard Model Physics}

		   GENIUS will allow besides the major step in neutrino 
		   physics described above the access to a broad range 
		   of other beyond SM physics topics in the multi-TeV range. 
	Already now $\beta\beta$ decay probes the TeV scale on which new 
	physics should manifest itself (see, e.g.% 
\cite{KK-Bey97,KK-LeptBar98,KK-SprTracts00}). 
	  Basing to a large extent on the theoretical work of the 
	  Heidelberg group in the last five years, the 
	  HEIDELBERG-MOSCOW experiment yields results for SUSY models 
	  (R-parity breaking, neutrino mass), leptoquarks 
	  (leptoquarks-Higgs coupling), compositeness, right-handed $W$ mass, 
	  nonconservation of Lorentz invariance and 
	  equivalence principle, mass of a heavy left or 
	  righthanded neutrino, competitive to corresponding results 
	  from high-energy accelerators like TEVATRON and HERA. 
	  The potential of GENIUS extends into the multi-TeV region for 
	  these fields and its sensitivity would correspond to that of 
	  LHC or NLC and beyond (for details see% 
\cite{KK60Y,KK-LeptBar98,KK-SprTracts00}).

%%%%%%%%%%%%%%%%%%%%%%%%%%%%% begin. fig. 8 %%%%%%%%%%%%%%%
\vspace{.75cm}
\begin{figure}[b]
%\vspace{9pt}
\centering{
\includegraphics*[scale=0.25]
{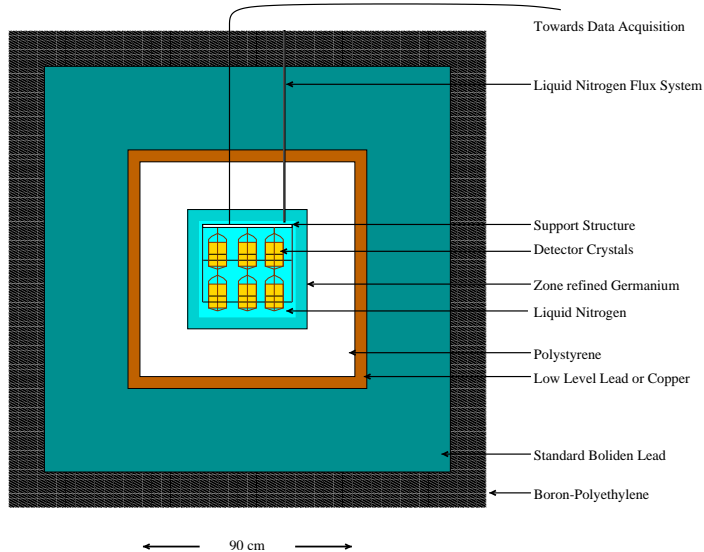}}

%\vspace{-.5cm}
\caption[]{%Figure 8. 
       Conceptual design of the Genius TF. Up
to 14 detectors will be housed in the inner detector chamber, filled with
liquid nitrogen. As a first shield 5 cm of zone refined Germanium will be used.
Behind the 20 cm of polystyrene isolation another 35 cm of low level lead
and a 15 cm borated polyethylene shield will complete the setup.
\label{fig:genius_tf_scheme}}
\end{figure}

%%%%%%%%%%%%%%%%%%%%%%%%%%%%%%%%%%%% end fig. 8 %%%%%%%%%%%%%%%%%%

%%%%%%%%%%%%%%%%%%% sect. 7 %%%%%%%%%%%%%%%%%%%%%%%%%%%%%

\section{GENIUS-Test Facility}
		     Construction of a test facility for 
		     GENIUS --- GENIUS-TF --- 
		     consisting of $\sim$~40 kg of HP Ge 
		     detectors suspended in a liquid nitrogen 
		     box has been started. 
		     Up to end of January 2001, four detectors each 
		     of $\sim$~2.5 kg and with a threshold of as low 
		     as $\sim$~500 eV have been produced.

         Besides test of various parameters of the GENIUS project, 
	 the test facility would allow, with the projected background 
	 of 2--4 events/(kg y keV) in the low-energy range, to probe the DAMA 
	 evidence for dark matter by the seasonal modulation signature 
	 within about one year of measurement with 95\% C.L.
	 Even for an initial lower mass of 20 kg the time scale would be 
	 not larger than three years, see (for details see% 
\cite{KK-GeTF-MPI,GenTF-0012022}). 
	 If using the enriched $^{76}$Ge detectors of the HEIDELBERG-MOSCOW 
	 experiment in the GENIUS-TF setup, a background in the 
	 $0\nu\beta\beta$ region a factor 30 smaller than in the 
	 HEIDELBERG-MOSCOW experiment could be obtained, 
	 which would allow to test the effective Majorana neutrino mass 
	 down to 0.15 eV (90\% C.L.)% 
\cite{KK-GeTF-MPI,GenTF-0012022}.

%%%%%%%%%%%%%%%%%%%%%%%%%%%%% Beg. Table 1 %%%%%%%%%%%%%%%%%%

\clearpage
\begin{table*}
\caption{Some key numbers of future double beta decay experiments (and of 
	the {\sf HEIDELBERG-MOSCOW} experiment). Explanations: 
	${\nabla}$ - assuming the background of the present pilot project. 
	$\ast\ast$ - with matrix element from  
\protect\cite{StMutKK90}, 
\protect\cite{Tom91}, 
\protect\cite{Hax84}, 
\protect\cite{WuStKKChTs91}, 
\protect\cite{WuStKuKK92} (see Table II in 
\protect\cite{HM99}). 
	${\triangle}$ - this case shown 
	to demonstrate {\bf the ultimate limit} of such experiments. 
	For details see 
\protect\cite{KK60Y}.}
\label{table:1}
\newcommand{\m}{\hphantom{$-$}}
\newcommand{\cc}[1]{\multicolumn{1}{c}{#1}}
\renewcommand{\tabcolsep}{.15 pc} % enlarge column spacing
\renewcommand{\arraystretch}{.75} % enlarge line spacing
{\footnotesize
%{\normalsize
{  
\begin{tabular}[!h]{|c|c|c|c|c|c|c|c|}
%[!h]{|c|c|c|c|c|c|c|c|}
\hline
\hline
 &  &  &  & Assumed &  &  & \\
 &  &  &  & backgr. & $Run-$ & Results & \\
$\beta\beta$-- & & & Mass & $\dag$ events/ & $ning$ & limit for & 
${<}m_{\nu}{>}$ \\
$Isoto-$ & $Name$ & $Status$ & $(ton-$ & kg y keV, & Time  
& $0\nu\beta\beta$ & \\
pe & & & $nes)$ & $\ddag$ events/kg & (tonn. & half-life & ( eV )\\ 
& & & & y FWHM,  & years) & (years) & \\
& & & & $\ast$ events &  &  & \\
& & &  & /yFWHM &  &  & \\
\hline
\hline
 &  &  &  &  &  &  & \\
~${\bf ^{76}{Ge}}$ & {\bf HEIDEL-} & {\bf run-}  & 0.011 & $\dag$ 0.06 
& {\bf 37.24} & ${\bf 2.1\cdot{10}^{25}}$ & {\bf $<$ 0.34} $\ast\ast$\\
 & {\bf BERG}  & {\bf ning}  &  (enri-  &  &  {\bf kg y} &  {\bf 90$\%$ c.l.} 
& {\bf 90$\%$ c.l.} \\
& {\bf MOSCOW} & {\bf since} & ched) & $\ddag$ 0.24  &  
& ${\bf 3.5\cdot{10}^{25}}$ & {\bf $<$ 0.26} $\ast\ast$\\
& {\bf \cite{KK-SprTracts00}} & {\bf 1990} &  & $\ast$ 2 & & 
{\bf 68$\%$ c.l.} & {\bf 68$\%$ c.l.}\\
& {\bf \cite{AnnRepGrSs00,KK01}} &  &  &  &  & {\bf NOW !!}  & {\bf NOW !!}\\
\hline
\hline
\hline
 &  &  &  &  &  &  & \\
${\bf ^{100}{Mo}}$ & {\sf NEMO III} & {\it under} & $\sim$0.01 & $\dag$ 
{\bf 0.0005} &  &  &\\
 & {\tt \cite{NEMO-Neutr00}}& {\it constr.} & (enri- & $\ddag$ 0.2  & 50 & 
${10}^{24}$ & 0.3-0.7\\
 &  & {\it end 2001?} & -ched) &  $\ast$ 2 &kg y  &  &\\
\hline
\hline
&  &  &  &  &  &  & \\
${\bf ^{130}{Te}}$ & ${\sf CUORE}^{\nabla}$ & {\it idea} & 0.75 & $\dag$ 0.5 
& 5 & $9\cdot{10}^{24}$ & 0.2-0.5\\
 & {\tt \cite{CUORE-LeptBar98}}& {\it since 1998} &(natural)  
& $\ddag$ 4.5/$\ast$ 1000 &  & & \\ 
\hline
&  &  &  &  &  &  &  \\
${\bf ^{130}{Te}}$ & {\sf CUORE}  &  {\it idea} & 0.75   & $\dag$ 0.005 & 5 
& $9\cdot{10}^{25}$ & 0.07-0.2\\
&  {\tt \cite{CUORE-LeptBar98,Fior-Neutr00}} & {\it since 1998}   & (natural)  
&  $\ddag$ 0.045/ $\ast$ 45 &  & &\\
\hline
&  &  &  &  &  &  & \\
${\bf ^{100}{Mo}}$ & {\sf MOON} & {\it idea} & 10 (enrich.) & ? & 30 & ? &\\
 & {\tt \cite{Ej00,LowNu2}} & {\it since 1999} &  100(nat.) & & 300 & &0.03 \\
\hline
&  &  &  &  &  &  & \\
${\bf ^{116}{Cd}}$ & {\sf CAMEOII} & {\it idea}  & 0.65 & * 3. & 5-8  
& ${10}^{26}$ & 0.06 \\
& {\sf CAMEOIII}{\sf \cite{Bell00}} & {\it since 2000 } & 1(enr.) & ? & 5-8 
&  ${10}^{27}$ & 0.02 \\
\hline
&  &  &  &  &  &  & \\
${\bf ^{136}{Xe}}$ & {\sf EXO} & Proposal& 1 & $\ast$ 0.4 & 5 & 
$8.3\cdot{10}^{26}$ & 0.05-0.14\\
&  & since &  &  &  &  & \\
  & {\tt \cite{EXO00,EXO-LowNu2}} & 1999 & 10 & $\ast$ 0.6 & 10 & 
$1.3\cdot{10}^{28}$ & 0.01-0.04\\
\hline 
\hline
\hline
\hline
&  &  &  &  &  &  &  \\
~${\bf ^{76}{Ge}}$ & {\bf GENIUS} & {\it under} & 11 kg & 
$\dag$ ${\bf 6\cdot{10}^{-3}}$& 3 
& {\bf ${\bf 1.6\cdot{10}^{26}}$} & {\bf 0.15} \\
& {\bf - TF} & {\it constr.} & (enr.) &  &  &  &   \\
&  {\bf \cite{KK-GeTF-MPI,GenTF-0012022}}&  {\it end 2001?} &  & &  &  &  \\
\hline
&  &  &  &  &  &  &  \\
~${\bf ^{76}{Ge}}$ & {\bf GENIUS} & Pro- & 1  & $\dag$ 
${\bf 0.04\cdot{10}^{-3}}$ & 1 & ${\bf 5.8\cdot{10}^{27}}$ & 
{\bf 0.02-0.05} \\
 & {\tt \cite{KK-Bey97,GEN-prop}}  & posal &(enrich.)  
& $\ddag$ ${\bf 0.15\cdot{10}^{-3}}$ & & & \\
&  & since &  & $\ast$ {\bf 0.15} &  &  &  \\
&  & 1997 & 1 & ${\bf \ast~ 1.5}$ & 10 & ${\bf 2\cdot{10}^{28}}$  & 
{\bf 0.01-0.028} \\
\hline
&  &  &  &  &  &  &  \\
~${\bf ^{76}{Ge}}$ & {\bf GENIUS} & Pro- & 10 
& $\ddag$ ${\bf 0.15\cdot{10}^{-3}}$ & 10 &
${\bf 6\cdot{10}^{28}}$ & {\bf 0.006 -}\\
&  {\tt \cite{KK-Bey97,GEN-prop}} &  posal &  &  &  &  &  {\bf 0.016}\\
 &   &  since &(enrich.) & ${\bf 0^{\triangle}}$ & 10 & 
${\bf 5.7\cdot{10}^{29}}$ & {\bf 0.002 -}\\
&  &  1997 &  &  &  &  &  {\bf 0.0056}\\ 
\hline 
\hline
\end{tabular}\\[2pt]
}}
%\vspace{0.3cm}
\end{table*}

%%%%%%%%%%%%%%%%%%%%%%%%%%%%%%%%%%%%%% end TABLE 1 %%%%%%%%%%%%%%%%%%%%%%%%
\clearpage
\noindent
This limit is similar to what some much larger experiments aim at 
(Table~1).

%%%%%%%%%%%%%%%%%%% end  sect. 7 %%%%%%%%%%%%%%%%%%%%%%%%%%%%%

%%%%%%%%%%%%%%%%%% tabl. 2 %%%%%%%%%%%%%%%%%%%%%%%%%%%
\begin{table*}[h]
\caption{Some of the new projects under discussion for future double beta 
	decay experiments (see ref.%
\protect\cite{KK60Y}).}
\label{tableA}
\newcommand{\m}{\hphantom{$-$}}
\newcommand{\cc}[1]{\multicolumn{1}{c}{#1}}
\renewcommand{\tabcolsep}{.19 pc} % enlarge column spacing
\renewcommand{\arraystretch}{.15} % enlarge line spacing
\begin{tabular}[!h]{|c|c|c|c|c|}
\hline
\hline
\multicolumn{5}{|c|}{}\\
\multicolumn{5}{|c|}{\Large NEW~~~  PROJECTS}\\
\multicolumn{5}{|c|}{}\\
\hline
 &  &  &  & \\
 & BACKGROUND & MASS & POTENTIAL & POTENTIAL \\
 &  &  &  & \\
 & REDUCTION & INCREASE & FOR DARK & FOR SOLAR\\
 &  &  & MATTER & ${\nu}^{'}$ s\\
\hline 
\hline
&  &  &  & \\
 {\sf GENIUS} 	& {\Large\bf +} 	& {\Large\bf +} 	
& {\Large\bf +} & {\Large\bf + ${}^{*)}$	}\\
\hline
&  &  &  & \\
 {\sf CUORE}  	& {\Large\bf (+)} 	& {\Large\bf + } 	
& {\Large\bf $-$} & {\Large\bf $-$	}	\\
\hline
&  &  &  & \\
 {\sf MOON}  		& {\Large\bf (+) }	& {\Large\bf + } 	
& {\Large\bf $-$} & {\Large\bf +}		\\
\hline
&  &  &  & \\
 {\sf EXO} 		&  {\Large\bf +} 	& {\Large\bf +}  	
& {\Large\bf $-$} & {\Large\bf $-$	}	\\
\hline
&  &  &  & \\
 {\sf MAJORANA}	& {\Large\bf $-$ }	& {\Large\bf + } 	
& {\Large\bf $-$} & {\Large\bf $-$	}	\\
\hline
\multicolumn{5}{|c|}{}\\
\multicolumn{5}{|c|}{\Large\sf *) real time measurement of pp neutrinos}\\
\multicolumn{5}{|c|}{\Large\sf with threshold of 10 keV (!!)}\\
\multicolumn{5}{|c|}{}\\
\hline
\hline
\end{tabular}\\[1pt]
%\vspace{0.3cm}
\end{table*}

%%%%%%%%%%%%%%%%%%end tabl. 2 %%%%%%%%%%%%%%%%%%%%%%%%%%%

%%%%%%%%%%%%%%%%%% conclusions %%%%%%%%%%%%%%%%%%%%%%%%%%%%%%
\vspace{-0.3cm}
\section{Conclusion}

	Nature is extremely generous to us, that with an increase of 
	the sensitivity by two orders of magnitude compared to the 
	present limit, down to 
$\langle m_\nu\rangle < 10^{-3}$~eV, 
	 indeed essentially all neutrino scenarios allowed by present 
	 neutrino oscillation experiments can be probed by double beta 
	decay experiments. 

	 GENIUS is the only of the new projects (see also 
Table ~2) 
	which exploits double beta decay to this ultimate limit.
	It is also the only one which simultaneously has a huge potential for 
      cold dark matter search, and for real-time detection of 
      low-energy neutrinos (see% 
\cite{KK-Bey97,Val01,AnnRepGrSs00,BedKK-01,KK-J-PhysG98,KK-NANPino00,KK-LowNu2}).

%%%%%%%%%%%%%%%%%%%%%%%%%%%%%%%%%%%%THE BIBLIOGR>%%%%%%%%%%%%%%%%%
\small{
        
}

\end{document}